\documentclass[aps,prd,twocolumn,english]{revtex4-1}
\usepackage[latin9]{inputenc}
\usepackage{amsmath}
\usepackage{amssymb}
\usepackage{graphicx}
\usepackage{babel}
\usepackage{color}
\usepackage[breaklinks = true,colorlinks = true,linkcolor = blue,citecolor = blue]{hyperref}

\newcommand{\CP}{\mathcal{CP}}

\newcommand{\calP}{\mathcal{P}}
\newcommand{\calT}{\mathcal{T}}
\newcommand{\bbar}{\bar{b}}
\newcommand{\boldb}{\boldsymbol{b}}
\newcommand{\bnabla}{\boldsymbol{\nabla}}
\newcommand{\bE}{\boldsymbol{E}}
\newcommand{\bB}{\boldsymbol{B}}
\newcommand{\bj}{\boldsymbol{j}}
\newcommand{\bk}{\boldsymbol{k}}
\newcommand{\calL}{\mathcal{L}}
\newcommand{\calE}{\mathcal{E}}
\newcommand{\Det}{\mathop{\mathrm{Det}}}

\usepackage[usenames,dvipsnames,svgnames,table]{xcolor}

\begin{document}
\title{Anomalous Casimir effect in axion electrodynamics}
\author{Kenji Fukushima}
\affiliation{Department of Physics, The University of Tokyo, %
             7-3-1 Hongo, Bunkyo-ku, Tokyo 113-0033, Japan}
\affiliation{Institute for Physics of Intelligence (IPI), The University of Tokyo, %
             7-3-1 Hongo, Bunkyo-ku, Tokyo 113-0033, Japan}
\author{Shota Imaki}
\affiliation{Department of Physics, The University of Tokyo, %
             7-3-1 Hongo, Bunkyo-ku, Tokyo 113-0033, Japan}
\author{Zebin Qiu}
\affiliation{Department of Physics, The University of Tokyo, %
             7-3-1 Hongo, Bunkyo-ku, Tokyo 113-0033, Japan}

\begin{abstract}
We study the Casimir effect in axion electrodynamics.
A finite $\theta$-term affects the energy dispersion relation of photon if $\theta$ is time and/or space dependent.
We focus on a special case with linearly inhomogeneous $\theta$ along the $z$-axis.
Then we demonstrate that the Casimir force between two parallel plates perpendicular to the $z$-axis can be either attractive or repulsive, dependent on the gradient of $\theta$.
We call this repulsive component in the Casimir force induced by inhomogeneous $\theta$ the anomalous Casimir effect.
\end{abstract}
\maketitle

\section{Introduction}

Casimir effect~\cite{Casimir:1948dh} refers to a physical force resulting from the quantum fluctuations in the vacuum restricted by boundaries.
One can alternatively interpret it as a relativistic extension of the
van~der~Waals force mediated by the vacuum polarization~\cite{Jaffe:2005vp}.
In the original study of the Casimir effect, an attractive force
between two parallel plates of perfect conductors emerges from the vacuum of Maxwell electrodynamics.
The presence of boundaries discretizes the momenta, causing
a finite difference in the vacuum energy, which is mathematically
represented by the Abel-Plana formula in the simple case with parallel plates.
The calculation machinery is quite analogous to the
imaginary-time formalism of finite-temperature field theory (see
Ref.~\cite{FUKUSHIMA2001455} for discussions on temperature inversion symmetry in the Casimir effect).
A pioneering experimental test for the Casimir force in the original scenario started out more than a half century ago~\cite{Sparnaay:1958wg}, while the more accurate measurement is established after decades of development
~\cite{Lamoreaux:1996wh,Mohideen:1998iz, Bressi:2002fr,Decca:2003zz}.
Theoretical generalizations of the original study include the dynamical Casimir effect
~\cite{Davies:1976hi,Calucci:1992hj,MaiaNeto:1993zz,Lambrecht:1996un,Golestanian:1997ks} and the fermionic Casimir effect
~\cite{Milton:1980uu,Elizalde:1997hx,Bulgac:2001np,Sundberg:2003tc}.
More interdisciplinarily, in view of modern nuclear and high-energy physics,
the Casimir effect shows great significance in the chiral bag model of
hadrons~\cite{Milton:1980ke,Milton:1982iw,Zahed:1984jq},
serves as a hypothetical candidate for the dark energy origin from
QCD~\cite{Urban:2009vy,Zhitnitsky:2011aa,Thomas:2012ib},
and also has important relevance to the researches of strings, branes, and gravity~\cite{Kikkawa:1984cp,Fabinger:2000jd,Puthoff:1989zz}.
Remarkably, the latest numerical simulations study the Casimir effect in Yang-Mills theory and relate it to nonperturbative mass
generation~\cite{Chernodub:2018pmt}.
Apart from such theoretical interests, moreover, the Casimir effect has considerable applications in the manufacture of micro electromechanical systems in nanotechnology~\cite{doi:10.1063/1.368410,PhysRevB.63.033402,Chan:2001zza,Chan:2001zzb,Genet2008}.

Among various aspects of the study on the Casimir effect, one intriguing issue is the sign of the Casimir force.
In fact, it has been demonstrated that the Casimir force can be flipped from attractive to repulsive via non-trivial geometry of
the boundaries~\cite{Boyer:1968uf,Mamaev:1979um,Levin:2010zz}.
The sign flip of the Casimir effect may also be caused by special
arrangements of objects and media with different permittivity or
permeability
~\cite{PhysRevA.9.2078,Alves:1999uh,Kenneth:2002ij,Rosa:2008zza,Zhao:2009zz}.
Interestingly, gathering substantial related efforts have given rise to a famous ``no-go'' theorem: the Casimir force between two bodies with reflection symmetry is always attractive~\cite{Kenneth:2006vr}.

However, this no-go theorem can be circumvented in consideration that
the ``vacuum'' in quantum field theory is not always trivial, but can have rich structures.
For such theories, even if boundaries maintain reflection symmetry, non-trivial vacuum properties may produce a repulsion.
Indeed, it has been argued that the sign flip of the Casimir force exists in the vacuum of the chiral Gross-Neveu model~\cite{Flachi:2017cdo}
or its scalar cousin, i.e., the $CP^{N-1}$ model~\cite{Flachi:2017xat}.
It is worth mentioning that these consequences have been numerically validated by first-principle simulations of lattice field theory in Ref.~\cite{Chernodub:2019nct}.
Lately, in a simpler setup even without interaction effects, a tunable Casimir force that can oscillate between attractive and repulsive has been derived by Ref.~\cite{Jiang:2018ivv}.
In their scenario, the key point lies in that the chiral media, i.e., optically active or gyrotropic media, endow the vacuum with an intrinsic breaking of spatial parity $\calP$ and/or time reversal $\calT$ symmetries.
In this way, without breaking the reflection symmetry geometrically, a repulsive Casimir force is allowed, which indicates a specific mechanism to bypass the no-go theorem.
Actually, materials exhibiting such intrinsic $\calP$ and/or $\calT$ symmetries breaking are familiar in condensed matter physics,
and for example, Refs.~\cite{Grushin:2010qoi,Rodriguez-Lopez:2013pza,Tse:2012pb,Wilson:2015wsa} have reported the repulsive Casimir force between two topological insulators, Hall materials, and Weyl semimetals, respectively.

Along these lines, we are pursuing a kindred mechanism for a repulsive
Casimir force in axion electrodynamics aka Maxwell-Chern-Simons theory~\cite{Carroll:1989vb,Jackiw:1999yp,PerezVictoria:1999uh}.
The Lagrangian density of axion electrodynamics contains an ordinary
electromagnetic part and a (3+1)-dimensional topological term
parameterized by background $\theta(x)$ which can be
interpreted as background axion field.
We note that our work should be distinguished from those in (2+1)-dimensional Chern-Simons electrodynamics~\cite{Milton:1990yj,Milton:1992rf,Bordag:1999ux,Alves:2010zzc}.
Given that a constant $\theta(x)$ would not affect the equation of motion, we consider a linearly inhomogeneous background axion field; $\theta(x) = b_\mu x^\mu$ with a constant four-vector
$b_\mu$.
This specific choice is also motivated by related works about the realization of quantum anomaly in condensed matter physics as discussed
in Refs.~\cite{Zyuzin:2012tv,Zyuzin:2012vn,Chen:2013mea,PhysRevLett.111.027201}
where a similar form of $\theta(x)$ is assumed.
Axion electrodynamics is a useful theory to account for anomaly induced
phenomena in chiral media, e.g., the Witten effect~\cite{Witten:1979ey}
the chiral magnetic
effect~\cite{Vilenkin:1980fu,Fukushima:2008xe,Fukushima:2018grm},
the anomalous Hall effect~\cite{PhysRev.95.1154,Jungwirth:2002zz,Haldane:2004zz,RevModPhys.82.1539}, etc.
Nowadays, Weyl semimetal, topological insulator, and axion crystal
provide us with real-world playgrounds of axion electrodynamics,
promoting cross-disciplinary studies and experimental searches for
anomalous chiral phenomena~\cite{Ooguri:2011aa,
Basar:2013iaa,Sekine:2015eaa,Ozaki:2016vwu,Barnes:2016fno,
Li:2016vlc,Tatsumi:2018ifx}.
In addition to activities in condensed matter physics,
one can see recent reviews~\cite{Kharzeev:2015znc,Shi:2017cpu,Zhao:2018skm}
for applications of chiral transport phenomena in the high-energy nuclear experiments.

There are preceding efforts on the Casimir effect in the
framework of axion electrodynamics.  In Ref.~\cite{Cao:2017ocv} the
topological Casimir effect was proposed as a possible probe to detect
the background $\theta$ angle and the QCD axion.  Thereby, a mixing
coupling between electric and magnetic fields plays an important role,
which is often referred to as magnetoelectric effect in condensed
matter physics.
More relevant to our present study is the work of
Ref.~\cite{Kharlanov:2009pv} where the Casimir effect with a pure timelike $b_\mu=(b_0,\boldsymbol{0})$ was analyzed, leading to the conclusion that no repulsive Casimir force is found in that case.
Our present work is a natural extension to the situation with a pure spacelike $b_\mu = (0, \boldb)$, and as we would argue later, we discover a repulsive component of Casimir force.

For the above-mentioned purpose, we carry out an analytical computation of the zero-point oscillation energy and the associated Casimir force in the presence of $\theta(x)$.
We adopt the original straightforward method by Casimir~\cite{Casimir:1948dh}
as well as some technical implementation similar to the case with
$b_\mu=(b_0,\boldsymbol{0})$ in Ref.~\cite{Kharlanov:2009pv}.
In Appendix, we provide calculations using alternative methodology based
on scattering theory and the Lifshitz formula~\cite{Lifshitz:1956zz},
which looks superficially different from Casimir's method but yields
an equivalent result.

The structure of this paper is organized as follows.
In Sec.~\ref{sec:axion} we present the definition of the theory we are interested in
and the physical setup to idealize the anomalous Casimir
effect in the presence of $b_\mu=(0,\boldb)$.
We proceed to concrete calculations of the vacuum energy in
Sec.~\ref{sec:vacuum}.  We figure out the energy dispersion relations
and quantify the zero-point oscillation energy there.
Section~\ref{sec:casimir} is devoted to our central results, i.e., the
analytical expression of the anomalous Casimir effect and the
numerical plot illustrating a repulsive region.  Finally, we conclude our discussions in Sec.~\ref{sec:conclusion}.

\section{Axion Electrodynamics}
\label{sec:axion}

We briefly introduce the axion electrodynamics and then expound our
physical scenario for a repulsive Casimir force.  We model the effect
of chiral medium on the Casimir force using the axion electrodynamics,
that is, the U(1) electrodynamics with a topological $\theta$ term
defined by the following Lagrangian density,
\begin{equation}
  \calL_{\text{axion}} =-\frac{1}{4} F_{\mu\nu} F^{\mu\nu}
  +\frac{1}{4} \theta F_{\mu\nu} \tilde{F}^{\mu\nu} \,.
  \label{eq:mcs_lagrangian}
\end{equation}
In the above expression
$F_{\mu\nu} \equiv \partial_\mu A_\nu - \partial_\nu A_\mu$ and
$\tilde{F}^{\mu\nu} \equiv \frac{1}{2} \epsilon^{\mu\nu\alpha\beta} F_{\alpha\beta}$,
where $A_\mu$ represents the U(1) gauge field.
For space-time dependent $\theta(x)$, the topological $\theta$ term
modifies the equations of motion.  Here, for later convenience, let us
denote its derivatives as
\begin{equation}
  b_0(x)
	\equiv \partial_t \theta(x)\,,\qquad
  \boldb(x) \equiv - \bnabla \theta(x)\,,
\end{equation}
or equivalently $b_{\mu}(x)=\partial_{\mu}\theta(x)$ in the covariant
notation.  Nonzero $b_0$ and/or $\boldb$ add $\CP$-odd terms to the
equations of motion.  The Euler-Lagrange equations from
Eq.~\eqref{eq:mcs_lagrangian},
\begin{equation}
\partial_{\mu}F^{\mu\nu}=b_{\mu}\tilde{F}^{\mu\nu}\,,
\end{equation}
and the Bianchi identity, $\partial_{\mu}\tilde{F}^{\mu\nu}=0$,
comprise the Maxwell-Chern-Simons equations in the absence of source.
The explicit forms read~\cite{Wilczek:1987mv}:
\begin{align}
  & \bnabla\cdot\bE = -\boldb\cdot\bB\,,
  \label{eq:mcs_charge}\\
 & \bnabla\times\bB-\frac{\partial\bE}{\partial t} =
  b_{0}\bB+\boldb\times\bE,\,
  \label{eq:mcs_current}\\
 & \bnabla\cdot\bB = 0\,,\\
 & \bnabla\times\bE+\frac{\partial\bB}{\partial t} = 0\,.
\end{align}
The first equation~\eqref{eq:mcs_charge} implies an extra charge
$-\boldb\cdot\bB$, which is commonly called the Witten
effect~\cite{Witten:1979ey}.  We can regard the right-hand side of
Eq.~\eqref{eq:mcs_current} together with Maxwell's displacement
current from the left-hand side as the current source for the magnetic
field.  Then, we find an extra current term,
$\bj_{\text{CME}}=b_{0}\bB$, which can be understood as
the chiral magnetic effect with the identification of $b_{0}$ as the chiral
chemical potential $\mu_{5}$.  Another extra current,
$\bj_{\text{AHE}}=\boldb\times\bE$, represents the anomalous
Hall effect which exists even without the magnetic field.

\begin{figure}
  \centering
\includegraphics[width=0.8\columnwidth]{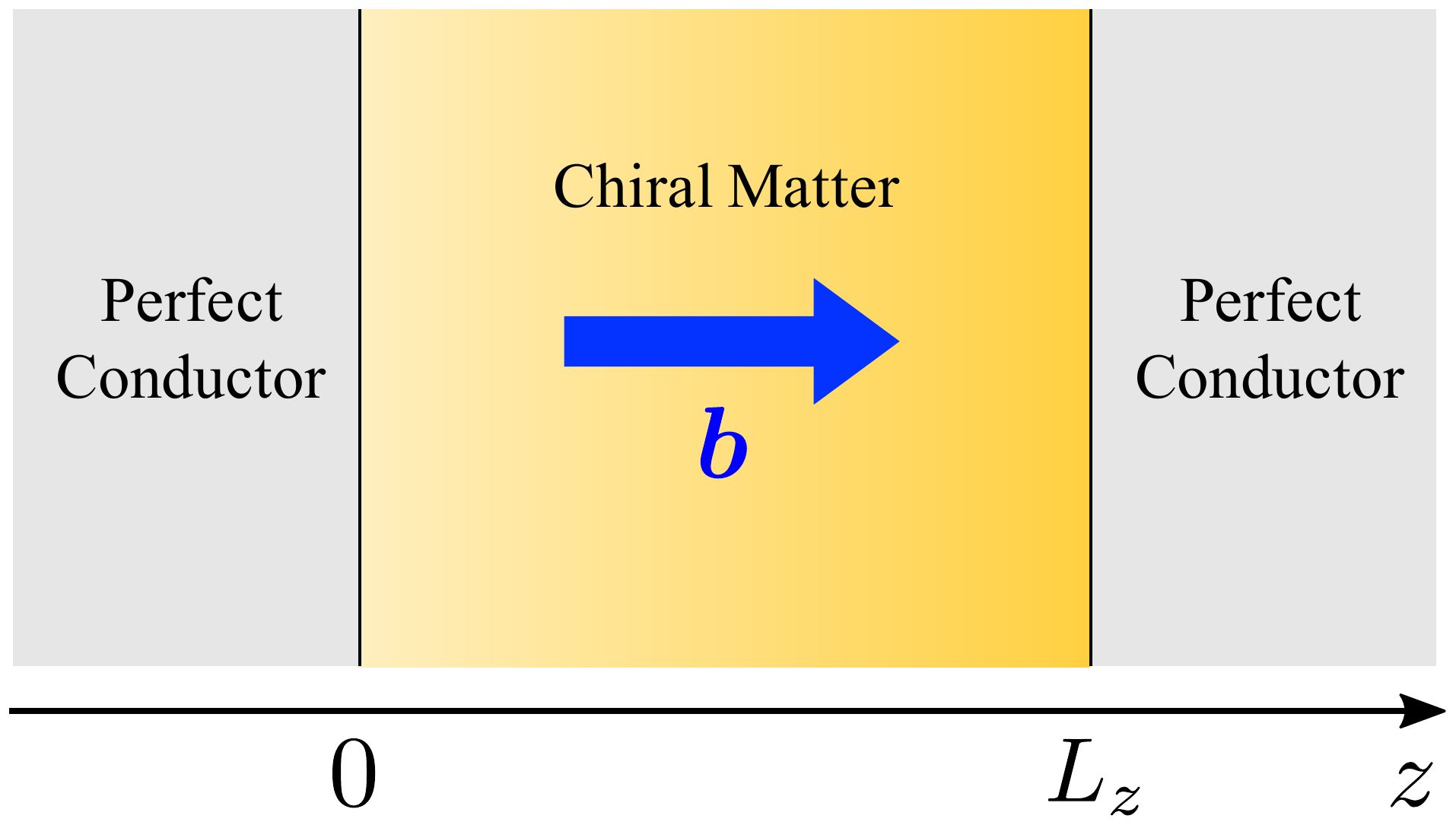}
\caption{Schematic illustration for the physical setup.
Two perfect conductor plates at $z=0$ and $z=L_{z}$ constitute the
transverse planes coordinated by $\hat{x}$ and $\hat{y}$.  The space
between two plates is filled with chiral matter represented by the
axion electrodynamics.}
\label{fig:scenario}
\end{figure}

These anomalous charge and currents are induced by the $\CP$-violating
modifications on the vacuum in the axion electrodynamics.  Since the
vacuum properties are such changed, we can naturally anticipate
noticeable impacts on other physical observables related to them.
In this work, specifically, we explore such possibility in terms of the
Casimir force. For such a purpose, as illustrated in
Fig.~\ref{fig:scenario}, we install two plates of perfect conductors
parallel to each other upright to the $z$-axis.  The interval distance
between two plates is $L_z$ and the size of each transverse plate is
$L_{x}L_{y}$.  For simplicity, we assume constant $b_0$ and $\boldb$.

It is known that a timelike $b_{\mu}$ may incur tachyonic
instabilities at long wavelength, which would impede the covariant
quantization of the electromagnetic
fields~\cite{Andrianov:1998wj,Kostelecky:2000mm,Adam:2001ma}.
Also, we point out that the Casimir effect with constant
$b_0\neq 0$ but $\boldb=0$ has been addressed in
Ref.~\cite{Kharlanov:2009pv}, where no sign flip of the Casimir force
was observed.  Thus, we focus on the situation with $b_0=0$ and
$\boldb\neq 0$ in the present work. For transverse symmetry,
we postulate $\boldb = b \hat{z}$, that is, $\boldb$ is directed
perpendicular to the two plates. In our setup with such
$\boldb\neq 0$, the reflection symmetry is explicitly broken, which
suggests that there may arise a repulsive component in the Casimir
force.  Indeed, we will confirm this with concrete calculations.


\section{Vacuum Energy}
\label{sec:vacuum}

We impose the Dirichlet boundary condition, $A_\mu = 0$, at $z = 0$ and
$z = L_z$, which is consistent with the properties of perfect
conductors.  Moreover, we take the limit $L_{x,y} \to \infty$.
Then, we discretize the electromagnetic wave vector as
$\bk = (k_x, k_y, k_z = n \pi / L_z)$ with $n \in \mathbb{Z}$.

A canonical quantization scheme for $A_\mu$ with covariant gauge was proposed in Ref.~\cite{Alfaro:2006dd,Colladay:2014dua,Colladay:2016rsf},
in which a tiny photon mass was introduced.
Instead, here we adopt
a path integral quantization with ghost fields.
The Lagrangian density with the gauge fixing term parameterized by $\xi$, and the ghost fields $c$ and $\bar{c}$, reads:
\begin{align}
  \mathcal{L}
	&= \mathcal{L}_{\text{photon}} + \mathcal{L}_{\text{ghost}} \notag \\
  &= -\frac{1}{4} F_{\mu\nu} F^{\mu\nu} - \frac{1}{4} b A_{\nu} \tilde{F}^{z \nu}
  + \frac{1}{2\xi} \left( \partial_\mu A^\mu \right)^{2}
  + \frac{1}{2} \partial_\mu \bar{c} \partial^\mu c \,.
\end{align}
The vacuum energy density $\varepsilon$ is obtained from the
generating functionals as follows:
\begin{equation}
  V T \varepsilon 
  = i \log Z_{\text{photon}} + i \log Z_{\text{ghost}}\,.
\end{equation}
Here $V = L_x L_y L_z$ is the volume of the vacuum region between
two plates and $T$ is the time interval in the path integral.
We keep them finite in the intermediate calculations and take the limits
of $L_{x,y} \to \infty$ and $T \to \infty$ in the end.
Beginning with the calculation of the photon part,
we rewrite the photon Lagrangian as a bilinear form of $A_\mu$ in momentum space:
\begin{equation}
	\calL_{\text{photon}}
	= - \frac{1}{2} A_\mu G_{\mu\nu}^{-1} A_\nu \,,
\end{equation}
where
\begin{equation}
  G_{\mu\nu}^{-1}
	= g_{\mu\nu} k^2 + i \epsilon_{\mu\nu\alpha\beta} b^\alpha k^\beta
	- \biggl( 1 - \frac{1}{\xi} \biggr) k_\mu k_\nu \,.
\end{equation}
Then we have:
\begin{equation}
 i \log Z_{\text{photon}}
 = -\frac{i}{2} \log \text{Det} \left[G_{\mu\nu}^{-1} (k) \right],
\end{equation}
where Det represents the determinant with respect to the momentum
index $k$ and the Lorentz indices $\mu$, $\nu$.
We firstly calculate the determinant over Lorentz indices as
\begin{equation}
  \Det \left[G_{\mu\nu}^{-1}(k)\right]
  = \prod_k \,
  \xi^{-1} \left(k^2\right)^2 \left[-\left(k^{2}\right)^2 + b^2 (k^2 + k_z^2)\right]\,.
\label{eq:DetG}
\end{equation}
For further calculations, we employ the following notation for the
energy dispersion relations determined from the on-shell condition~\cite{Qiu:2016hzd}:
\begin{align}
	\omega_{1,2}^2 &= \bk^2 \,, \\
	\omega_{\pm}^2 &= k_x^2 + k_y^2 + \left( \sqrt{k_z^2 + \frac{b^2}{4}} \pm \frac{b}{2} \right)^2 \,.
\label{eq:dispersion}
\end{align}
We note that $\omega_{1,2}$ are zeros of $(k^2)^2$ in Eq.~\eqref{eq:DetG}.
Since $(k^2)^2$ appears from the longitudinal and the scalar polarizations, the modes with $\omega_{1,2}$ are unphysical and
their contributions to vacuum energy are canceled by the ghosts.
The physical modes $\omega_\pm$ are
zeros of $-(k^2)^2+b^2 (k^2+k_z^2)$ and they correspond to the right- and left-handed photons. With these dispersion relations,
we express the vacuum energy contributed from the photon as
\begin{equation}
  i \log Z_{\text{photon}}
	= - \sum_{i = 1, 2, \pm} \sum_k \,
  \frac{i}{2} \log \left[k_0^2 - \omega_i^2 (\boldsymbol{k}) \right],
\label{eq:Z_photon}
\end{equation}
where we have
dropped an irrelevant constant $\xi^{-1}$.
By a similar computation for the ghost, we acquire:
\begin{equation}
	i \log Z_{\text{ghost}}
	= 2 \sum_k \frac{i}{2} \log \left(k_0^2 - \boldsymbol{k}^2 \right) \,.
	\label{eq:Z_ghost}
\end{equation}
Notably Eq.~\eqref{eq:Z_ghost} cancels the contribution
from the unphysical modes with $i=1,2$ in Eq.~\eqref{eq:Z_photon}.
Summing the photon and the ghost contributions up,
we get,
\begin{equation}
  V T \varepsilon
	= - \frac{i}{2} \sum_{\pm} \sum_k
  \log \left[k_0^2 - \omega_\pm^2 (\boldsymbol{k})\right] \,.
\end{equation}
Now, we take the limits of $L_x, L_y, T \to \infty$, which replace
the phase space sum over $k = (k_0, k_x, k_y, n \pi / L_z)$ as
\begin{equation}
  \frac{1}{VT} \sum_k \; \to \;
  \frac{1}{L_z} \sum_{n = 0}^\infty \int \frac{dk_0 dk_x dk_y}{(2 \pi)^3}\,.
\end{equation}
Here, let us briefly explain how to compute the $k_0$-integral.
Differentiating the integral
with respect to $\omega_\pm$, we find two poles on the real $k_0$ axis.
We deform the poles by the standard $i \epsilon$ prescription and carry out the $k_0$-integration.
After further integrating over $\omega_\pm$,
we extract a finite $\omega_\pm$-dependent piece, dropping
an irrelevant divergent part: 
\begin{equation}
  \int \frac{d k_0}{2 \pi} \log \left(k_0^2 -\omega_\pm^2 + i \epsilon \right)
  = i \omega_\pm + \text{(const.)} \,.
  \label{eq:k0int}
\end{equation}
Eventually,
we attain the vacuum energy per unit transverse area,
$\mathcal{E} = V \varepsilon / L_x L_y$, given by
\begin{equation}
  \calE
	= \sum_{\pm} \sum_{n = 0}^\infty
  \int \frac{dk_x dk_y}{(2 \pi)^2} \, \frac{\omega_\pm (\bk)}{2} \,.
  \label{eq:E}
\end{equation}
Equation~\eqref{eq:E} is nothing but the sum of the zero-point oscillation energy,
sharing the same structure as the
conventional Casimir effect except for the energy dispersion relations.

\section{Casimir Force}
\label{sec:casimir}

Based on the vacuum energy
achieved in the last section,
we will quantify
the Casimir
force in this section.
We bring in a new notation to express the energy dispersion relation:
\begin{equation}
  \omega_\pm^2(\bk)
	= k_x^2 + k_y^2 + \frac{\pi^2}{L_z^2} \mu_\pm^2(n)\,,
\end{equation}
where
\begin{equation}
	\mu_\pm (n)
	\equiv \sqrt{n^2 + \bbar^2} \pm \bbar\,.
\end{equation}
Here, $\bbar$ denotes the dimensionless $b$ defined by
$\bbar = bL_z/(2\pi)$.
We rescale
$k_{x, y} \to \tilde{k}_{x,y} \equiv (L_z / \pi \mu_\pm ) k_{x, y}$,
to perform the transverse momentum integration as
\begin{align}
  \calE
	&= \frac{\pi^3}{L_z^3}
	\sum_{\pm} \sum_{n = 0}^\infty \,
	\mu_\pm^3 (n)
	\int^{\tilde{\Lambda}_\pm} \frac{d \tilde{k}_x d \tilde{k}_y}{(2 \pi)^2} \;
	\frac{1}{2} \sqrt{\tilde{k}_x^2 + \tilde{k}_y^2 + 1} \,,
	\label{eq:energyImu}
\end{align}
where we introduced an ultraviolet cutoff $\Lambda_\pm$ so that a step function, $\Theta(\Lambda_\pm-\omega_\pm(\bk))$, is convoluted in the
integrand.  With the rescaled dimensionless
cutoff,
$\tilde{\Lambda}_\pm\equiv (L_z/\pi\mu_\pm)\Lambda_\pm$,
we further evaluate the integral in Eq.~\eqref{eq:energyImu} as
\begin{align}
	& \int^{\tilde{\Lambda}_\pm} \frac{d \tilde k_x d \tilde k_y}{(2 \pi)^2}
	\sqrt{ \tilde k_x^2 + \tilde k_x^2 + 1 } \notag \\
	&= \frac{1}{2 \pi} \int_0^\infty
	dk_r\, k_r\, \sqrt{k_r^2 + 1}\,
	\Theta(\tilde{\Lambda}_\pm - \sqrt{k_r^2 + 1}) \, \notag \\
	&= -\frac{1}{6 \pi} + \frac{\tilde{\Lambda}_\pm^3}{6 \pi} \,.
	\label{eq:Ipm}
\end{align}
After inserting Eq.~\eqref{eq:Ipm} into Eq.~\eqref{eq:energyImu},
the second term proportional to $\tilde{\Lambda}_\pm^3$ 
leads to an irrelevant constant independent of $L_z$.
We therefore safely leave this term out and reduce the expression of $\calE$ to
\begin{align}
  \calE
		&= -\frac{\pi^2}{12 L_z^3} \sum_{\pm} \sum_{n=0}^\infty \mu_\pm (n)^3
	\label{eq:Ebyn}
\end{align}
Taking the sum over $\pm$, we further simplify the above expression
into:
\begin{equation}
	\calE
	= - \frac{\pi^2}{12 L_z^3} \biggl[ S \Bigl(-\frac{3}{2}, \bbar\Bigr)
	+ 3 \bbar^2 S\Bigl(-\frac{1}{2}, \bbar\Bigr)\biggr]
	- \frac{b^3}{24 \pi}\,,
	\label{eq:EbyS}
\end{equation}
where we defined a function,
\begin{align}
	S(s,\bbar)
	&\equiv \sum_{n = -\infty}^\infty \left(n^2 + \bbar^2\right)^{-s} \,,
	\label{eq:S}
\end{align}
for which we note that the sum runs from $n = -\infty$ to $\infty$.
The last term in Eq.~\eqref{eq:EbyS}, coming from $n = 0$,
is independent of $L_z$ and hence gives no contribution to the Casimir force.
Therefore we safely drop this term hereafter.
Thus, our problem boils down to the calculation of
$S(s,\bbar)$.
We rewrite Eq.~\eqref{eq:S} by multiplying it with the integral form of
$\Gamma(s)$ and then divide by $\Gamma(s)$ as
\begin{equation}
	S(s,\bbar)
	= \sum_{n = -\infty}^{\infty} \frac{1}{\Gamma (s)}
	\int_0^\infty \left(n^2 + \bbar^2\right)^{-s} u^{s-1} e^{-u} du \,.
\end{equation}
We change the integration variable from $u$ to $v=u/(n^2+\bbar^2)$
so that the integral becomes
\begin{align}
	S(s,\bbar)
	&= \sum_{n = -\infty}^{\infty} \frac{1}{\Gamma (s)}
	\int_0^\infty v^{s-1} e^{-n^2 v-\bbar^2 v} dv \notag \\
	&= \sum_{m=-\infty}^\infty \frac{\sqrt{\pi}}{\Gamma(s)}
	\int_0^\infty v^{s-3/2} e^{-\pi^2 m^2/v -\bbar^2 v} dv \,,
	\label{eq:S}
\end{align}
where we used Poisson's summation formula.
The term with $m = 0$ yields the Gamma function, while
the terms with $m \neq 0$ take the form of the integral representation
for the modified Bessel function of the second kind. 
We, therefore, arrive at
\begin{equation}
	S(s,\bbar)
	=\frac{\sqrt{\pi} \bbar^{1-2s}}{\Gamma(s)}
	\left[
	\Gamma \left(s - \frac{1}{2} \right)
	+ 4 \sum_{m = 1}^\infty \frac{K_{\frac{1}{2} - s} (2 \pi m \bbar)}{(\pi m \bbar)^{\frac{1}{2} - s}}
	\right] \,.
\end{equation}
Plugging this to the Casimir energy \eqref{eq:EbyS}, we get
\begin{equation}
	\begin{split}
		\calE
		&= \calE_\infty + \calE_{\text{reg}} \notag\\
                &= \calE_\infty
		+ \frac{b^4 L_z}{16 \pi^2}
		\sum_{m = 1}^\infty \left[
			\frac{K_1(m b L_z)}{m b L_z} -\frac{K_2(m b L_z)}{(m b L_z)^2}
		\right] \,.
	\end{split}
	\label{eq:EC}
\end{equation}
Here, the energy per unit transverse area, $\calE$, includes a divergent portion,
\begin{equation}
	\calE_\infty
	= -\frac{5 b^4 L_z}{512 \pi^3} \Gamma(0) \,.
\end{equation}
But the corresponding energy density $\calE_\infty / L_z$ is independent of $L_z$.
Thus, we can harmlessly subtract this energy density
irrelevant to the Casimir force, by shifting a reference level of the
energy density.

Finally, the Casimir force per unit transverse area is given by the derivative of $\calE_{\text{reg}}$ with respect to $L_z$, that is,
\begin{align}
	F(b)
	&= - \frac{\partial \calE_{\text{reg}}}{\partial L_z} \notag \\
	&= -\frac{b^4}{16 \pi^2}
	\sum_{m = 1}^\infty
	\left[ \frac{3 K_2 (m b L_z)}{(m b L_z)^2} - K_0 (m b L_z) \right] \,.
\label{eq:F}
\end{align}
This is our central result.
We note that the limiting behaviors
$K_2(x) \to 2x^{-2}$ and $K_0(x) \to \log x$ for $x \to 0$ result in
\begin{equation}
	F(0)
	\equiv \lim_{b \to 0} F(b)
	= - \frac{3}{8 \pi^2 L_z^4} \sum_{m = 1}^\infty \frac{1}{m^4}
	= - \frac{\pi^2}{240 L_z^4} \,,
\end{equation}
which retrieves the well-known result within the Maxwell electrodynamics.

\begin{figure}[]
\centering
\includegraphics[width=\columnwidth]{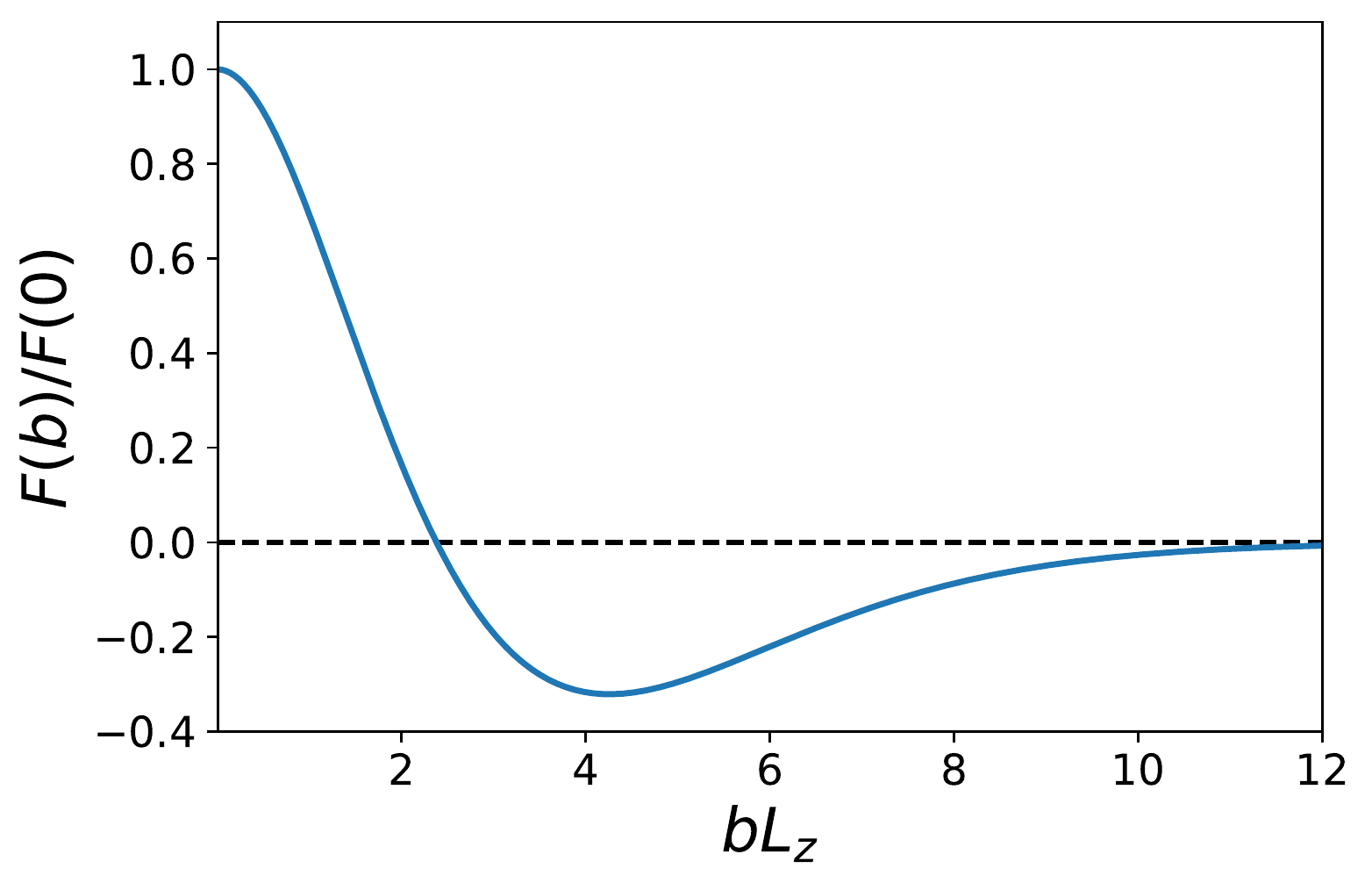}
\caption{Casimir force as a function of the dimensionless distance
  scaled with $b$.}
\label{fig:graph}
\end{figure}

The $b$-dependence of the Casimir force \eqref{eq:F} is shown in Fig.~\ref{fig:graph}.
One can observe that the Casimir force is repulsive when $b L_z > 2.38$.
By tuning the distance between two plates while keeping $b L_z$ larger than $2.38$,
the strength of the repulsive Casimir force is, in principle, arbitrarily tunable.
The ratio $F(b) / F(0)$ takes the minimum value $-0.32$ for $b L_z = 4.26.$
In the physical units, this extremal value of repulsive force is
estimated as $3.95 \times 10^{-5} (b^4 [\mu{\rm m}^4]) {\rm dyn}/{\rm cm^2}$.
We note that our results qualitatively match Ref.~\cite{Jiang:2018ivv}
for $b L_z \ll 1$.
In Appendix, we present an alternative approach developed in
Ref.~\cite{Jiang:2018ivv} to reproduce exactly the same numerical
result of $F(b)$ as in Fig.~\ref{fig:graph}. Such an
independent calculation based on different subtraction procedures
serves as a double check for our results and a confirmation for our
scheme to subtract infinities in
Eqs.~\eqref{eq:k0int}, \eqref{eq:Ipm}, and \eqref{eq:EC}.


\section{Conclusion}
\label{sec:conclusion}

We demonstrated a repulsive component of the Casimir force in axion electrodynamics by formulating its explicit expression in an analytically closed form.
We circumvented the no-go theorem which tends to forbid the repulsive Casimir force between two objects with reflection symmetry.
Our underlying idea consists in the intrinsic parity symmetry breaking in the chiral vacuum between the plates, which is quite analogous
to a recent proposal in Ref.~\cite{Jiang:2018ivv}.

Our next step is to seek for experimental realization of our
theoretical consequence.
Our physical setup, in which the $\theta$-angle has a spatial gradient
perpendicular to plates, would be realized through topological materials.
For instance, a Weyl semimetal with the separation between Weyl nodes features the gradient of the $\theta$-angle in the electromagnetic effective action.
Besides, it has been proposed that the periodically-stacked structure of trivial and topological insulators
also generates the gradient of the $\theta$-angle~\cite{Ozaki:2016vwu}.
Another promising proposal to engender the gradient of the
$\theta$-angle is to utilize an external rotating electric field supplied by a circularly polarized laser to irradiate Dirac semimetal~\cite{Ebihara:2015aca}.
These examples are feasible candidates for realizing the repulsive Casimir force revealed in the present paper.

\acknowledgments
K.~F.\ was supported by Japan Society for the Promotion
of Science (JSPS) KAKENHI Grant No. 18H01211.
S.~I.\ was supported by Grant-in-Aid for JSPS Fellows Grant Number 19J22323.

\appendix

\section{Scattering Formalism for the Casimir Force}
\label{sec:appendix}

We here supply an alternative methodology to derive
the result~\eqref{eq:F}, referring to the scattering theory approach developed
in Ref.~\cite{Jiang:2018ivv}. The scenario in
Ref.~\cite{Jiang:2018ivv} considers a situation of inserting
non-trivial electromagnetic material in between two perfect conductor
plates. Concretely speaking, their material features the birefringence
parameterized by a constant shift in the $z$-component of the wave vector, $\delta k_{z}$,
with the dispersion relation following from classical Maxwell electrodynamics,
i.e.,
\begin{equation}
\omega_{0}^{2}=k_{x}^{2}+k_{y}^{2}+\left(\bar{k}_{z}\pm\delta k_{z}\right)^{2}.\label{eq:dispersion_0}
\end{equation}
As demonstrated by Eq.~\eqref{eq:E}, all possible modifications on the Casimir
energy in the vacuum of axion electrodynamics are encoded in the non-trivial dispersion relation.
Then, despite physical distinction between the chirality origins in
their and our studies,
the formula derived in Ref.~\cite{Jiang:2018ivv} is directly
applicable for our current setup.
To this end, we just need to replace their dispersion
relation~\eqref{eq:dispersion_0} with ours in Eq.~\eqref{eq:dispersion}.

Hence, we quote their expression for the Casimir energy:
\begin{equation}
\calE_{\text{reg}}
= \int_0^\infty \frac{d \zeta}{2 \pi} \int_{-\infty}^\infty \frac{d k_x d k_y}{(2 \pi)^2}
\log\det \left(\mathbb{I} - R_1 U_{12} R_2 U_{21}\right) \,,
\label{eq:EC_scattering}
\end{equation}
where $\zeta=-i\omega$ is the imaginary frequency, $R_{1}$ and $R_{2}$
refer to the reflection matrix of plate 1 at $z=0$ and plate 2 at
$z=L_{z}$, respectively, and $U_{12}$ and $U_{21}$ stand for the
translation matrix from plate 1 to 2 and from 2 to 1.
We note that above $\calE_{\text{reg}}$ may have an $L_z$ independent
discrepancy from Eq.~\eqref{eq:EC}, which would make no
difference in the force.
For a chiral medium between two planes,
the translation matrices in helicity basis read:
\begin{equation}
U_{12}=\left(\begin{array}{cc}
e^{ik_{z}^{+}L_{z}} & 0\\
0 & e^{ik_{z}^{-}L_{z}}
\end{array}\right),\quad U_{21}=\left(\begin{array}{cc}
e^{ik_{z}^{-}L_{z}} & 0\\
0 & e^{ik_{z}^{+}L_{z}}
\end{array}\right),\label{eq:translation}
\end{equation}
in accordance with the plane wave ansatz in Euclidean geometry.
Here, the dispersion relations~\eqref{eq:dispersion} are expressed
with $k_{z}^{\pm}$ given in terms of $\zeta$ as
\begin{equation}
k_{z}^{\pm}\equiv i\sqrt{\zeta^{2}+\boldsymbol{k}_{\perp}^{2}\pm ib\sqrt{\zeta^{2}+\boldsymbol{k}_{\perp}^{2}}}\,.\label{eq:kz+-}
\end{equation}
Meanwhile, for two identical perfect conductor plates, the reflection
matrix can be ideally presumed as:
\begin{equation}
R_{1}=R_{2}=\left(\begin{array}{cc}
0 & -1\\
-1 & 0
\end{array}\right).\label{eq:reflection}
\end{equation}
We insert Eqs.~\eqref{eq:translation} to \eqref{eq:reflection} into Eq.~\eqref{eq:EC_scattering}
to obtain the Casimir energy $\calE_{\text{reg}}$. The spatial derivative
finally yields the Casimir force:
\begin{align}
F & =-\frac{d\calE_{\text{reg}}}{dL_{z}}=\int_{0}^{\infty}\frac{d\zeta}{2\pi}\int_{-\infty}^{\infty}\frac{d^{2}\boldsymbol{k}_{\perp}}{\left(2\pi\right)^{2}}\nonumber \\
 & \times\frac{1\!-\!2i\bigl(k_{z}^{+}e^{2ik_{z}^{+}L_{z}}\!+\!k_{z}^{-}e^{2ik_{z}^{-}L_{z}}\bigr)\!+\!2i(k_{z}^{+}\!+\!k_{z}^{-})e^{2i(k_{z}^{+}+k_{z}^{-})L_{z}}}{1-e^{2ik_{z}^{+}L_{z}}-e^{2ik_{z}^{-}L_{z}}+e^{2i\left(k_{z}^{+}+k_{z}^{-}\right)L_{z}}}.\label{eq:F_scattering}
\end{align}
It can be proved that Eq.~\eqref{eq:F_scattering} is equivalent to Eq.~\eqref{eq:F}, and the numerical plot matches Fig.~\ref{fig:graph} perfectly.

\bibliography{Casimir}{}
\bibliographystyle{apsrev4-1}

\end{document}